\documentclass[11pt, a4paper]{article}
\pdfoutput=1
\usepackage{jheppub}
\usepackage{multirow}
\usepackage{slashed}
\usepackage{float}
\usepackage[caption=false]{subfig}
\usepackage{tabularx}
\usepackage[normalem]{ulem}

\newcommand{\stau}{\tilde{\tau}_1}

\newcolumntype{L}[1]{>{\raggedright\arraybackslash}p{#1}}
\newcolumntype{C}[1]{>{\centering\arraybackslash}p{#1}}
\newcolumntype{R}[1]{>{\raggedleft\arraybackslash}p{#1}}

\preprint{LAPTH-016/16, HRI-RECAPP-2016-007}
                                                                          
\begin{document}
\title{Signatures of sneutrino dark matter in an extension of the CMSSM}
\author{Shankha Banerjee$^1$, Genevi\`{e}ve B\'{e}langer$^1$, Biswarup Mukhopadhyaya$^2$, Pasquale D. Serpico$^1$}
\affiliation[1]{LAPTH, Univ. de Savoie, CNRS, B.P.110, F-74941 Annecy-le-Vieux, France}
\affiliation[2]{Regional Centre for Accelerator-based Particle Physics, Harish-Chandra Research Institute,
Chhatnag Road, Jhusi, Allahabad 211019, India}

\abstract{Current data (LHC direct searches, Higgs mass, dark matter-related bounds) severely affect the constrained minimal SUSY standard model (CMSSM) with neutralinos as dark matter candidates. But the evidence for neutrino masses coming from oscillations requires extending the SM with at least right-handed neutrinos
with a Dirac mass term. In turn, this implies extending the CMSSM with right-handed sneutrino superpartners, a scenario we dub $\tilde\nu$CMSSM. These additional states
constitute alternative dark matter candidates of the superWIMP type, produced via the decay of the long-lived next-to-lightest SUSY particle (NLSP). 
Here we consider the interesting and likely case where the NLSP is a $\tilde\tau$: despite the modest extension with respect 
to the CMSSM this scenario has the distinctive signatures of heavy, stable charged particles. After taking into 
account the role played by neutrino mass bounds and the specific cosmological bounds from the big bang nucleosynthesis in 
selecting the viable parameter space, we discuss the excellent discovery prospects for this model at the future runs of the LHC. We show that it is possible to probe $\stau$ masses up to 600 GeV at the
14 TeV LHC with $\mathcal{L} = 1100$ fb$^{-1}$ when one considers a pair production of staus with two or more hard jets through
all SUSY processes. We also show the complementary discovery prospects from a direct $\stau$ pair production, as well as at the new experiment MoEDAL.}

\maketitle
\section{Introduction}\label{sec:1}

The search for supersymmetry (SUSY) broken around the TeV scale has been so far unsuccessful at the Large Hadron Collider (LHC)~\cite{Aad:2016jxj,Khachatryan:2016kdk,Aad:2015baa}. This, together with the requirement of having a Higgs boson mass around 125 GeV~\cite{Aad:2012tfa,Chatrchyan:2012xdj}, puts strong pressure on the idea of SUSY as a solution to the naturalness problem~\cite{Barbieri:1987fn,deCarlos:1993rbr,Hall:2011aa},
a situation further exacerbated if one requires the theory to provide a  dark matter candidate matching the relic abundance, now determined to exquisite precision~\cite{Ade:2015xua}.
It is no surprise that the simplest (read most economical) version of SUSY theories becomes the first casualty: the constrained minimal SUSY standard model (CMSSM), based on minimal supergravity (mSUGRA)~\cite{Chamseddine:1982jx} is currently strongly disfavoured~\cite{Baer:2012mv,Ghilencea:2012gz}. This is due to the rigid links existing in the different sectors of the theory: direct collider bounds push the strongly interacting sector (squarks and gluinos) to heavy scales, but this also translates into heavy sleptons and gauginos, since they are all largely controlled by the universal gaugino ($m_{1/2}$) and scalar ($m_0$) masses. This not only portrays a dismal picture for collider searches and the naturalness argument, but also spells trouble for the scenario where dark matter (DM) is made of (dominantly bino) neutralinos, since the heavy mass scales suppression of the dominant annihilation cross-sections generically leads to a too large relic abundance.
 In addition, the option of a higgsino-like DM candidate face constraints from direct searches~\cite{Akerib:2015rjg} unless its mass is at the TeV scale. The latter faces naturalness issues since the  Higgsino mass parameter 
$\mu$ is determined in terms of  scalar masses in this framework, via the electroweak symmetry breaking condition. Such inter-connected nature of the CMSSM parameters, 
 which all  evolve from a limited set ($m_0$,  $m_{1/2}$, $A_0$, $\tan\beta$ and the sign of $\mu$), 
enhances the difficulty in finding a suitable DM candidate.
 Coupled with the fact that $m_0$  and  $m_{1/2}$ also affect the Higgs mass(es), all this has caused 
the  CMSSM to run into rough weather~\cite{Strege:2012bt,Kim:2013uxa,Buchmueller:2013rsa,Roszkowski:2014wqa,Bechtle:2015nua}.

The situation, however, can be quite different with a `minor' change in the particle 
spectrum, which is in fact suggested by an empirical argument. It is known that the MSSM has no built-in mechanism for generating neutrino masses, 
the evidence for which has grown in the past two decades into an inescapable reality~\cite{Ahmad:2002jz,Fukuda:1998mi,Ahn:2002up}, for which the 2015 Nobel Prize in Physics has been recently awarded. The most immediate solution to amend the MSSM
is to add three right-handed (RH) neutrino superfields, and the corresponding terms
in the superpotential, which  would lead to their Yukawa interactions. This also
implies the simultaneous existence of three right-chiral sneutrinos, one of  which 
could now  be the lightest SUSY particle (LSP) and act as potential DM 
candidate. However, at least for Dirac neutrino masses,  
the Yukawa interactions are extremely small ($\simeq 10^{-13}$), and
any interaction of the right-sneutrino  is proportional 
to this coupling.
As a result, the sneutrino LSP never reaches thermal equilibrium, its abundance being determined by the decay of heavier weakly interacting massive particles (WIMPs)
rather than by its annihilation into SM particles as in the standard WIMP freeze-out picture.
At the same time, the highly suppressed interaction strength automatically allows such a  kind of DM candidate to evade the limits coming from direct DM search
experiments in underground detectors. The lack of tight restrictions on the DM mass can in principle relax the
constraints on the superparticle spectrum in the CMSSM. 
Hence in such a scenario one can expect different allowed regions in the parameter space than in the standard CMSSM, with 
correspondingly peculiar cosmological constraints and observable signals at the LHC. The present
work is devoted to an investigation in these directions. The collider signatures of this model will differ significantly from the cases where the sneutrino can be a thermal DM either because of a large mixing with other chiral sneutrino  states~\cite{Arina:2007tm,Arina:2015uea,Kakizaki:2015nua,Dumont:2012ee,Belanger:2011ny,Arina:2013zca,Thomas:2007bu,BhupalDev:2012ru,Banerjee:2013fga}, or because the sneutrino couples to some new particle~\cite{Belanger:2011rs,Belanger:2015cra,Cerdeno:2009dv}.

It is obvious that the lightest SUSY particle belonging to the CMSSM spectrum, which we shall also loosely call next-to-lightest SUSY particle (NLSP)~\footnote{Within the states available in the $\tilde\nu$CMSSM, accounting for the three generation structure, strictly speaking the NLSP is in fact the second lightest sneutrino, almost degenerate with the LSP in what follows.}  is long-lived if its
only $R$-parity conserving~\footnote{The multiplicative conservation of 
$R = (-1)^{(3B+L+S)}$---with $B, L,$ and $S$ denoting the baryon number, the lepton number, and the spin of the particle, respectively---ensures that the LSP is stable and a DM candidate.} decay into the LSP is driven by the tiny neutrino Yukawa
coupling. 
In scenarios, where the MSSM scalar masses evolve from a universal  $m_0$, the NLSP will generally be a neutralino or the lighter stau.     Adding the RH sneutrinos will allow to relax the astrophysical and cosmological constraints on scenarios with a neutralino NLSP,  however the collider signatures being similar to those of the standard CMSSM we will not consider this class of scenarios. We will rather concentrate on the case where the stau  is the lightest sparticle in the MSSM spectrum - a scenario that can only be made viable with the presence of a sneutrino LSP (or a gravitino). Note that the  gravitino mass is an arbitrary parameter in the CMSSM, the gravitino could therefore  be the LSP, we will not consider this possibility any further as it   has been studied   in ~\cite{Feng:2003xh,Feng:2004mt,Ellis:2003dn,Baer:2014eja}. 
Contrary to the elusive LSP, it should be noted that the NLSP should satisfy several cosmological constraints. First, since
it is a WIMP-like progenitor of the LSP, its thermal relic abundance obtained  
upon solving the Boltzmann equation for ${\tilde \tau}_1$  is subject to standard constraints, modulo the rescaling by $m_{{\tilde \nu}_R}/m_{{\tilde \tau}_1}$. 
Secondly, as will be discussed below in detail, the late decay of the stau NLSP may tamper
with light element abundances and jeopardise the big bang nucleosynthesis
(BBN) predictions. This puts upper limits on its abundance vs. lifetime, leading to constraints
on its mass and interaction strengths. 

A stau-NLSP that decays into the sneutrino DM candidate consistent with the
observed neutrino mass scale will be stable on the distance scale of a collider detector. Therefore, contrary to customary SUSY scenarios with large missing-$E_T$ signatures, 
these models are characterized by a pair of highly ionising charged tracks that are seen both in the inner tracking chamber
and the muon chamber. This has prompted efforts to identify such tracks via
time-delay measurements, leading to lower limits on the order of 350 GeV on 
the masses of stable staus of this kind.  In an earlier publication ~\cite{Gupta:2007ui}, some of us have shown that
certain kinematical selection criteria are particularly effective in separating signal and background, especially for relatively low masses of 
the stable charged particles. These  criteria include cuts on the track transverse momenta ($p_T$), the scalar sum 
of the  $p_T$'s of all visible particles, and the invariant mass of the pair
of the two most highly ionising tracks in any event. It was also demonstrated
in \cite{Biswas:2009rba,Biswas:2009zp} that events selected with the help of these criteria could be used to reconstruct certain superparticle masses. 
Collider signatures with charged tracks are also expected in models with a substantially
massive gravitino LSP~\cite{Heisig:2011dr,Heisig:2012zq} or  with almost degenerate $\tilde\tau_1$-neutralino LSP~\cite{Heisig:2015yla}.
In Ref.~\cite{Heisig:2013rya}, the reach for the stau NLSP has been studied in the context of the pMSSM.

Compared with earlier investigations~\cite{Ishiwata:2009gs}, this paper presents a number of improvements: 
first, we update the cosmological constraints which  not only include the relic abundance, but  also
the upper limit on the neutrino mass and, as argued, the BBN constraints.
At the same time, we impose LHC null searches performed till now as well as the requirement of obtaining
a scalar mass around 125 GeV, in a scenario where the MSSM spectrum follows
from the CMSSM postulates.
 The allowed spectra thus found are subjected to our proposed selection criteria for
the 14 TeV run. We identify in this manner (a) the currently viable $\tilde\nu$CMSSM
parameter space with  ${\tilde \tau}_1$-NLSP, ${\tilde \nu}_R$-LSP, and (b) the
regions that can be probed at the LHC with gradually accumulating luminosity.

In section~\ref{ModelConstraints} we briefly describe the extended CMSSM model and list all the constraints, \textit{viz.}
constraints from the relic abundance, from the elemental 
abundance of $^4$He and $^2$H and the existing constraints from runs I and II at
the LHC. We review the constraints from Big Bang Nucleosynthesis in section~\ref{BBN}. We show the existing available parameter
space after implementing all the constraints in section~\ref{Results}. In section~\ref{LHC14}, we discuss few prospective channels
through which the available parameter space can be probed via LHC14 in the High luminosity run at 3000 fb$^{-1}$. Finally we 
summarise and conclude in section~\ref{conclusions}, where we also discuss some possible directions for future studies.

\section{Model and constraints}
In this section we start by discussing the framework and then we will summarise all the existing constraints used in this 
analysis. Here we consider the MSSM augmented with three generations of RH (s)neutrinos, with a Dirac mass term (implying extremely small Yukawa couplings). This 
$\tilde\nu$CMSSM model is the simplest extension of the CMSSM which can explain non-zero masses and mixing of the neutrinos.
In the present work we consider lepton number conservation, hence the MSSM superpotential is extended by just one term
for each family,
\begin{equation}
 W_{\nu}^R = y_{\nu} \hat{H}_u \hat{L} \hat{\nu}^c_R,
\end{equation}
where $y_{\nu}$ is the neutrino Yukawa coupling, $\hat{L}=(\hat{\nu}_L,\hat{\ell}_{\bar{L}})$ is the left-handed (LH) lepton 
superfield, $\hat{H}_u = (\hat{H}^+_u, \hat{H}^0_u)$ is the Higgs superfield which gives masses to the $T_3 = +1/2$ fermions and 
$\hat{\nu}_R$ is the superfield for the RH neutrinos. This superpotential ensures the presence of RH
sneutrinos in the particle spectrum. These sneutrinos will have all their couplings proportional to the corresponding 
neutrino masses.  We will consider mainly the case where  neutrinos are degenerate and the sneutrino mass term is universal, hence the sneutrinos will be nearly degenerate. We will assume that the lightest sneutrino that might become the LSP  is the RH eigenstate of tau-sneutrino.

In this model, after symmetry breaking the neutrinos obtain their masses as
\begin{equation}
 m_{\nu} = \frac{y_{\nu}}{\sqrt{2}} v \sin{\beta},
\end{equation}
where $v \simeq 246.2$ GeV is the vacuum expectation of the SM-like Higgs boson and 
$\tan{\beta} = \big< H_u^0 \big>/ \big< H_d^0 \big>$.
While the details of the sneutrino DM scenario described here are sensitive to the matrix structure of the Yukawa couplings,
almost all the qualitative features only rely on the smallness of the Yukawa, with their overall size being the key quantitative parameter, determining for instance the very small decay rate of the $\stau$-NLSP. We shall use the currently allowed range of the largest neutrino mass,  $m_{\nu}^H$,
as a proxy for the size of the relevant Yukawa coupling. A lower bound on the coupling can be inferred from global fits of the neutrino oscillation parameters to solar, atmospheric,
reactor and accelerator neutrino data, which provide at 3 $\sigma$ the range for the largest mass-squared splitting~\cite{Capozzi:2013csa},
\begin{equation}
 |\Delta m^2|\equiv |m_3^2 - (m_1^2 + m_2^2)/2|= 2.43 (2.38)\pm 0.06 \, \times 10^{-3} \, \textrm{eV}^2\,,
\end{equation}
where $m_i$ are the three neutrino masses and the number in parenthesis is for the inverted hierarchy scenario $(m_3 < m_1 < m_2)$. 
The heaviest mass ($ m_{\nu}^H$) is thus bounded by
\begin{equation}
 m_{\nu}^H \geq \sqrt{|\Delta m^2|} \simeq 0.049\, {\rm eV}\,,
\end{equation}
with the equality attained only for hierarchical neutrino masses, when it yields
\begin{equation}
 (y_{\nu}^H \sin{\beta})_{\rm min} \simeq 2.8 \times 10^{-13}\,.\label{osc1min}
\end{equation}
The upper limit on this Yukawa coupling follows instead from the upper limit on the absolute neutrino mass scale, which is currently dominated by the cosmological bound
on the sum of neutrino masses. The recent combination~\cite{Ade:2015xua} of Planck temperature  (TT) and polarisation (lowP) data with lensing and 
external data  including supernovae, Baryon acoustic oscillation (BAO) and the astrophysical determination of the Hubble constant $H_0$ yields (see~\cite{Ade:2015xua} for details)
\begin{equation}
 \sum_{i=1}^{3}{m_{ \rm i}} < 0.23 \; \textrm{eV} \, \textrm{at 95\% CL};  \,\label{sumPlanck1}
\end{equation}
This upper limit  translates---for a quasi-degenerate neutrino mass spectrum---into $m_{\nu}^H \lesssim 0.077\,$eV which implies 
\begin{equation}
 (y_{\nu}^H \sin{\beta} )_{\rm max}\simeq 4.4 \times 10^{-13}\,.\label{yukmax}
\end{equation}
One must note that this number depends to some extent on the number and type of datasets analysed and on the theoretical model assumed for the cosmological fit.
The upper limit Eq.~(\ref{sumPlanck1}) could be tightened by a factor $\sim$2 (see for instance~\cite{Cuesta:2015iho}), essentially leading the allowed Yukawa coupling interval to collapse to the value of Eq.~(\ref{osc1min}), or relaxed by a similar factor of $\sim 2\div 3$ (see~\cite{Ade:2015xua} for details). Given the fast progress in the field of cosmology, we shall present our ``fiducial'' results for the Yukawa coupling corresponding to Eq.~(\ref{yukmax}), but will also show the impact of lowering its value to the one of Eq.~(\ref{osc1min}).

In the CMSSM,  SUSY breaking is introduced by
universal soft terms for the scalars ($m_0$) and  the gauginos ($m_{1/2}$) along with the trilinear couplings $A_0$ and  the bilinear term for the Higgs, $B$,  in the Lagrangian at some high scale.
The $B$ parameter and the supersymmetric Higgs mass parameter, $\mu$  are  determined by the electroweak symmetry breaking conditions (up to the sign of $\mu$). 
Once the soft SUSY breaking parameters are specified at a high scale ($\mathcal{O}(10^{15})\,$GeV) and $\tan\beta$ is fixed at the electroweak scale  one can determine the masses of all  the squarks, sleptons, gauginos as well as  the mass parameters of the Higgs sector using the renormalization group equations (RG). 
In the $\tilde\nu$CMSSM, the RH sneutrino has little impact on the rest of the spectrum, hence the superparticle spectrum almost exactly mimics the one obtained in the CMSSM  save for the fact that now the LSP can be the RH sneutrino. 
Neglecting any inter-family mixing, the additional mass term for the sneutrinos reads

\begin{equation}
 -\mathcal{L}_{soft} \supset M_{\tilde{\nu}_R}^2 |\tilde{\nu}_R|^2 + (y_{\nu} A_{\nu} H_u \, \tilde{L}\, \tilde{\nu}_R^c + \, h.c.),
\end{equation}
where $A_{\nu}$ is responsible for the left-right mixing in the scalar mass matrix. It is obtained by the running of the trilinear
soft SUSY breaking term, $A_0$.  Note that we assume  a sneutrino mixing that depends on $y_{\nu} A_{\nu}$ as is typically the case 
in  SUGRA-inspired scenario where the trilinear soft terms arise from F-terms. From the RG equation solution it is shown in Ref.~\cite{Asaka:2005cn} that at the scale $m_Z$, $A_{\nu}$ is given by $ A_{\nu} = A_0 - 0.59 m_{1/2}$. 
The left-right mixing angle of the sneutrino  $\tilde\Theta$ is given by
\begin{equation}
 \tan{2\tilde\Theta} = \frac{2 y_{\nu} v \sin{\beta} |\cot{\beta} \mu - A_{\nu}|}{m_{\tilde{\nu}_L}^2 - m_{\tilde{\nu}_R}^2}\,.
 \label{eq:mixing}
\end{equation}
Owing to the fact the neutrino Yukawa couplings are extremely small, the sneutrino  mixing  can be neglected.
 We consider the LH and  RH sneutrinos as mass eigenstates with
\begin{equation}
 m_{\tilde{\nu}_L}^2 = M_{\tilde{L}}^2 + \frac{1}{2} m_Z^2 \cos{2\beta} \;\;\; \textrm{and} \;\;\; m_{\tilde{\nu}_R}^2 = M_{\tilde{\nu}_R}^2,
\end{equation}
where $M_{\tilde{L}}$ and $M_{\tilde{\nu}_R}$ are the soft scalar masses for the LH sleptons and the RH
neutrinos respectively. The right chiral neutrino superfield is different from the remaining fields in MSSM; 
it has no gauge interaction, and it interacts with MSSM fields only via the Yukawa terms in the superpotential, where again 
the interaction strengths are very different from those for the other fields, leading to extremely small neutrino masses. All 
this suggests a somewhat separate status for these superfields, including the possibility of its being actually a member of 
a hidden sector. In view of all this, one may like to accord a different origin for the right neutrino soft masses, as 
compared to those arising from $m_0$. Keeping this in mind, we will not necessarily
require $M_{\tilde{\nu}_R} = m_0$, although we do comment on the consequence of this assumption.

Note that the RG evolution of all the parameters of the CMSSM remain almost unaffected
in the $\tilde\nu$CMSSM, with the evolution of the new states being almost negligible:
The RH sneutrino mass parameter evolves at one-loop level 
as~\cite{Martin:1997ns}
\begin{equation}
 \frac{dM_{\tilde{\nu}_R}^2}{dt} = \frac{2}{16 \pi^2}y_{\nu}^2 A_{\nu}^2\,.
\end{equation}
Hence, the smallness of the Yukawa coupling ensures that $M_{\tilde{\nu}_R}$ remains basically fixed at its UV value,
whereas all the other sfermion masses evolve up at the electroweak scale.
It is also  worth noting that all  three right sneutrinos are similar in nature and, for a universal  value of the matrix $M_{\tilde{\nu}_R}$ eigenvalues at high scale,
 one has a near-degeneracy of three RH sneutrinos, with splittings $\delta M_{\tilde{\nu}_R}^2$ of the order of the neutrino mass splittings $\delta m_{\nu}^2$. Thus  the universal GUT scale conditions on the  parameters of an R-parity conserving scenario can generate a spectrum where the three 
RH sneutrinos will be stable (or metastable but very long-lived), leading to different decay chains
for supersymmetric particles as compared to those with a neutralino LSP.

To make the discussion more comprehensive, we discuss one important reason for choosing a stau-NLSP~\cite{Gupta:2007ui}. In general one can
also have a neutralino or a chargino NLSP as they are the remaining R-odd weakly interacting particles. However, a neutralino
NLSP will always end up decaying to a neutrino and a sneutrino leading to  a fully invisible final state. Hence
the collider signals will be almost exactly the same as for a model with a neutralino LSP.  A chargino NLSP 
can have different signatures through charged tracks. However, it is very difficult to have a model where the lighter
chargino ($\chi_1^{\pm}$) is lighter than the lightest neutralino ($\chi_1^0$), although it can happen at tree-level in specific corners of the MSSM~\cite{Kribs:2008hq}. On the other hand, it is very easy to accommodate a $\stau$-NLSP in the $\tilde\nu$CMSSM scenario, which
is thus the case we concentrate on in the following.

The $\stau$-NLSP eventually decays into the RH sneutrinos via $\stau \to W^{(*)} \tilde{\nu}_R$ (actually all $\tilde{\nu}_R$ states in the degenerate case
we are focusing on) driven by the tiny neutrino Yukawa coupling.
If we further assume $m_{\stau} > m_{\tilde{\nu}_R} + m_W$, the two-body decay width is given by~\cite{Asaka:2005cn}
\begin{equation}
 \Gamma_{\stau} \simeq \Gamma_{\stau \to \tilde{\nu}_R W} = \frac{g^2 \tilde\Theta^2}{32 \pi}|U_{L1}^{(\stau)}|^2 \frac{m_{\stau}^3}{m_W^2}\left[1 - \frac{2(m_{\tilde{\nu}_R}^2 + m_W^2)}{m_{\stau}^2} +  \frac{(m_{\tilde{\nu}_R}^2 - m_W^2)^2}{m_{\stau}^4}\right]^{3/2},
\end{equation}
where $g$ is the $SU(2)_L$ gauge coupling constant, $m_W$ is the mass of the $W$-boson and $U^{(\stau)}$ is the mixing matrix of
the staus which relate the two mass eigenstates  (here $m_{\stau} \leq m_{\tilde{\tau}_2}$) and the gauge eigenstates as
\begin{equation}
 \begin{pmatrix}
  \tilde{\tau}_L \\
  \tilde{\tau}_R
 \end{pmatrix} = U^{(\tilde{\tau})}
 \begin{pmatrix}
  \stau \\
  \tilde{\tau}_2
 \end{pmatrix}\,,
\end{equation}
and the subscript $L1$ indicates the $(1,1)^{\textrm{th}}$ element of this matrix.  When
$m_{\stau} < m_{\tilde{\nu}_R} + m_W$ the two-body decay is kinematically forbidden and the dominant three body
decays are $\stau \to \tilde{\nu}_R \ell \bar{\nu}, \tilde{\nu}_R q \bar{q}'$. The stau lifetime strongly depend on the decay modes and on the mixing in the $\tilde\nu$ and $\tilde\tau$ sectors, typical lifetimes range from a few seconds to over $10^{11}\,$s.

The lifetime of the NLSP is long  enough that its decay occurs well after its freeze-out, yet  it has been shown in Ref.~\cite{Asaka:2005cn,Asaka:2006fs} that the $\tilde{\nu}_R$  retains all good properties as cold DM, being in particular stable due to R-parity conservation, and evading direct detection constraints due to the suppressed interactions from the tiny Yukawa coupling. 
The density parameter of $\tilde{\nu}_R$ from the decay of the NLSP after freeze-out is simply given by
\begin{equation}
 \Omega_{\tilde{\nu}_R} = \frac{m_{\tilde{\nu}_R}}{m_{\stau}} \Omega_{\stau},
\end{equation}
where $\Omega_{\stau}$ is the present density parameter of the NLSP assumed stable. 
In the present work $\Omega_{\stau}$ is computed using the code {\tt micrOMEGAs}~\cite{Belanger:2001fz,Belanger:2014vza}.  Note that in the following we neglect any enhancement
in the DM abundance that could come from other production channels, such as heavier sleptons decays (either directly into $\tilde{\nu}_R$ or, more likely, via $\stau$). 
Especially for moderately degenerate slepton mass spectra, these could enhance $\Omega_{\tilde{\nu}_R}$ by an amount that can be estimated at the {\cal O}(10\%) level, which should be kept in mind. 
Recent cosmological data~\cite{Ade:2015xua} yield 
\begin{equation}
 \Omega_{\textrm{DM}} h^2 = 0.1199 \pm 0.0027\,.\label{cosmoODM}
\end{equation}

We shall use Eq.~(\ref{cosmoODM}) as a constraint, requiring that $\Omega_{\tilde{\nu}_R}< \Omega_{\rm DM}^{\rm max}$, for which
we use the 2 $\sigma $ upper value. In general, we shall find that  $\Omega_{\tilde{\nu}_R}< \Omega_{\rm DM}$, although in some
region of parameter space  $\Omega_{\tilde{\nu}_R}\simeq \Omega_{\rm DM}$, \textit{i.e}. the $\tilde{\nu}_R$ produced via $\stau$ decay
may constitute a sizable (if not dominant) constituent of the DM. Note that this can happen for a range of parameters 
 when the NLSP (which is actually the CMSSM-LSP) is charged, a situation very different  from the case where the neutralino constitutes the CMSSM-LSP.
A few more particle physics tools are used and constraints are imposed to ensure the phenomenological viability of our model:
\begin{itemize}
 \item The CMSSM spectrum is generated using  SPheno~\cite{Porod:2003um,Porod:2011nf}.
 \item The mass of the lightest Higgs is required to be in the range $123 \; \textrm{GeV} \; < m_{h^0} < 128 \; \textrm{GeV}$,  consistent with the Higgs mass measurements from the different channels at the LHC~\cite{Aad:2015zhl} after allowing for  a $\simeq 2$ GeV theoretical uncertainty.
 \item The signal strengths of the SM-like Higgs boson are required to match the experimentally quoted numbers~\cite{Aad:2015gba,Khachatryan:2014jba}. For this we
       use the package Lilith~\cite{Bernon:2015hsa} which computes the likelihood function and rejects parameter points 
       incompatible with the signal strength measurements.
 \item We demand that the mass of the long lived $\stau > 340$ GeV, which is the bound obtained by 
       CMS~\cite{CMS-PAS-EXO-15-010,Chatrchyan:2013oca} from the run I data for a direct pair production of staus. The bound from 
       ATLAS~\cite{ATLAS:2014fka} is slightly weaker, \textit{viz.} $\stau > 289$ GeV. 
 \item We further impose the $2\sigma$ bounds from  $b \to s \gamma$ at NLO~\cite{HFAG_bsg},  $B_s \to \mu^+ \mu^-$, ~\cite{CMS:2014xfa} and
       $\bar{B}^+ \to \tau^+ \nu_{\tau}$~\cite{HFAG_btau},  as computed with {\tt micrOMEGAs}~\cite{Belanger:2014vza}. 
 \item 
 We demand that $m_{\tilde{g}} > 1.8$ TeV,  this value correspond to the the limit obtained from the LHC Run II data~\cite{ATLAS-CONF-2015-067} that extends on the already stringent bound of Run 1
~\cite{Aad:2015iea}. Note that this bound refers to the CMSSM and does not apply directly to  the $\tilde\nu$CMSSM that we consider here. However, because of the relation between supersymmetric particles masses, the lower bound on the long-lived stau mentioned above forces the gluino to be rather heavy in our model.
        \item Finally we consider what are perhaps the most important constraints in this study, \textit{i.e.} the constraints on the light nuclei produced in BBN. This constraint is discussed in  details in section~\ref{BBN}.
\end{itemize}

\label{ModelConstraints}
\section{Bounds from Big Bang Nucleosynthesis}
\label{BBN}
Despite the fact that nuclear binding energies range in the ballpark of several MeV per nucleon, as long as the temperature $T\gg 0.1\,$ MeV, virtually no nuclear species is present in the early universe, since the high entropy conditions cause the immediate photo-destruction of any bound states that forms. Standard primordial nucleosynthesis (for reviews,
see for instance~\cite{Iocco:2008va,Cyburt:2015mya}) describes the departure from the early phase of nuclear statistical equilibrium until the synthesis of light nuclei in the
cooling plasma is completed, at $T\sim {\cal O}(10)\,$keV. Since all processes happen at the kinetic equilibrium, in standard BBN the energies available for the nuclear reactions are limited, and the process can be described in a relatively simple and robust way.

When long-lived states are present in the early universe, if their  decay injects energetic particles with {\it visible}~\footnote{This excludes dark byproducts and to a large extent neutrinos.} energy per decay $E_{vis} \gg T$, they can trigger complicated {\it non-thermal} nuclear processes (non-thermal BBN).  Qualitatively, there are two types of processes and constraints. In general, one always expects sizable fractions of energy to be injected in the form of  non-thermal photons and electrons (e.m. channels, henceforth simply ``photons''): the associated constraints are very strong for lifetimes exceeding $\sim 10^6\,$s, see~\cite{Poulin:2015woa} and notably~\cite{Poulin:2015opa} for a recent overview and
treatment of these processes including a regime overlooked so far. This constraint however excludes only the long-lifetime tail of the viable $\stau$'s parameter space,  weakening significantly for lower lifetimes and vanishing  for injection times below $\sim 10^4\,$s.
This is due to the fact that $e^{\pm}$ pair production by energetic photons onto the thermal bath ones is extremely efficient in cutting off the high-energy tail of the photon spectra, with a cutoff energy that increases with decreasing plasma temperature. The cutoff must be larger than the threshold for photodisintegration cross sections in order for these processes to be relevant, which implies efficient bounds only at sufficiently late times~\cite{Iocco:2008va}.

However, the  $\stau$'s in the bulk of the parameter space of interest  have lifetimes shorter than $10^4\,$s: for those, the relevant bounds are due 
to the {\it hadronic} part of the cascades (with branching ratio $B_h$) induced by the stau decays, notably via the effects of mesons in altering the weak $n \leftrightarrow p$ equilibrium and of non-thermal nucleons on the nuclear reactions, e.g. via spallations~\cite{Iocco:2008va}.
This dynamics can only be described properly via Monte Carlo simulations, see for instance~\cite{Kawasaki:2004qu,Jedamzik:2006xz}, since some energy-losses are intrinsically stochastic, and using averages in deterministic equations may be inaccurate.  Here we base our constraints on
the results obtained in~\cite{Kawasaki:2004qu}. However, we do take into account the newer determinations of the abundances of light
nuclei, notably $^2$H and $^4$He, for which sufficiently precise measurements exist and for which the primordial origin
of the bulk of their abundance is not disputed~\cite{Iocco:2008va,Cyburt:2015mya}.
As explicitly noted in~\cite{Jedamzik:2006xz} (see its Appendix A),  when a
single process dominates the production (or destruction) of a given nucleus, a good approximation consists in assuming a linear relation for the change in the number of nuclei with respect to  the standard BBN yield vs. the number of decays/particle injected (at any given time). This property is used in our analysis since this ``single process dominance'' is well satisfied in our scenarios. The $^4$He bound relies on its overproduction  due to the  alteration of the $n/p$ ratio in the early ($t \lesssim 10\,$s) BBN phase, with little to no role for the alteration of the nuclear network;  the deuterium bound comes essentially from requiring that  $^2$H is not overproduced via hadro-disintegration of $^4$He, see Table IV in~\cite{Kawasaki:2004qu}. 
Note also that the bounds reported in~\cite{Asaka:2006fs} and that we want to update  are based on the results of
Fig. 39 in~\cite{Kawasaki:2004qu}.  For our reference standard computations, we rely on the \texttt{PArthENoPE} code (see~\cite{Pisanti:2007hk,Iocco:2008va}), 
which for an adopted baryon to photon ratio of $\eta = 6.1 \times 10^{-10}$ yields best-fit predictions of $Y_p = 0.2463$ and
$^2$H/H = 2.578 $\times 10^{-5}$ for $^4$He mass fraction and deuterium number density, respectively. It has been recently checked~\cite{Cyburt:2015mya} that very similar values (at the permil level) are obtained  in an updated version of the  Kawano code~\cite{Kawano:1992ua}, hence we expect a good agreement of the above figures with the baseline values that had been adopted  in~\cite{Kawasaki:2004qu}, which relied on an updated Kawano code.

Between the two  curves for $^4$He reported in~\cite{Asaka:2006fs}, the most relevant one is the constraint relying on the determination~\cite{Izotov:2003xn}, denoted as IT.  In fact, if we take the 2$\,\sigma$ upper limit by summing in quadrature the statistical and systematic error as reported in Eq. (2.4) of~\cite{Kawasaki:2004qu}, 
the $^4$He change leading to the 95\% C.L. exclusion in~\cite{Kawasaki:2004qu} can be estimated as 
$\Delta Y_p = 0.0066$.  The recent determination reported in~\cite{Cyburt:2015mya} (see Eq. (7) at zero metallicity)
would lead to a two sigma upper limit $Y_p^{\textrm{max}} \simeq 0.2529$, implying {\it by accident} exactly the same maximal allowed variation used a 
decade ago! We conclude that the state-of-the art limit coming from  $^4$He coincides to a good extent, albeit serendipitously, with the €œIT curve quoted in~\cite{Kawasaki:2004qu}, which we shall use henceforth.

Concerning deuterium  we will consider the ``low abundance" bound presented in  Ref.~\cite{Kawasaki:2004qu}  as a reference point since, 
relying on the combination of measurements reported in~\cite{Kirkman:2003uv},  it uses an abundance
$^2$H/H = 2.78$^{+0.44}_{-0.38} \times 10^{-5}$  which  is quite close to a more recent determination. This corresponds
to an estimated maximal  allowed change $\Delta ^2$H/H $\simeq 1.08 \times 10^{-5}$. 
Based on modern observations, even the conservative ranges used now in the literature converge to a 
more restrictive variation, for instance $\Delta ^2$H/H$\lesssim  0.9 \times 10^{-5}$ according to~\cite{2012MNRAS.426.1427O}, 
or $\Delta ^2$H/H=$ 0.73 \times 10^{-5}$ based on the compilation in~\cite{Iocco:2008va}, i.e. bounds between 20\% and 50\% more stringent.
If considering the best measurement available  $\Delta ^2$H/H $= (2.53 \pm 0.04) \times 10^{-5}$~\cite{Cooke:2013cba}
(see also Eq. (8) in~\cite{Cyburt:2015mya}) as representative of the state of the art, the largest variation allowed could be as small as
$\Delta ^2$H/H$\simeq 0.03 \times 10^{-5}$ , \textit{i.e.} the room for an exotic effect would have shrank by a factor $\sim 36$ as compared with the  
old estimates! Despite this huge improvement, the impact of the  deuterium determination  on the  existing bounds on $\stau$'s is only moderate, since the
deuterium constraint is a very sharp function of the lifetime, behaving almost like a step function around lifetimes of $\sim$100 s.
 In the following sections, unless otherwise stated, we shall conservatively assume the allowed deuterium interval improved by a mere 20\% 
over the one corresponding to the old ``€œlow abundance'' determination of Fig. 39 of~\cite{Kawasaki:2004qu}, we call this the conservative constraint. Yet, in order to gauge the impact of possible much higher precision in deuterium observations, we shall also compare those bounds to the constraint coming from the effect of a tightening of the maximal allowed ``exotic'' deuterium production by a factor 36, corresponding to the most optimistic/aggressive current estimates.

\section{Results}\label{Results}

In this section we explore the parameter space of the $\tilde\nu$CMSSM satisfying all 
constraints listed in the above two sections. 
 The input parameters of the model  at GUT scale are varied in the range 
 \begin{equation}
 m_0 < 2500\, {\rm GeV} \;;\;\;\; m_{1/2} < 2500\, {\rm GeV}\;;\;\; |A_0| < 3000\,  {\rm GeV}\;;\;\;\; 
 \label{eq:range1}
 \end{equation}
 while at the electroweak scale 
 \begin{equation}
  0< m_{\tilde \nu_R}< m_{\stau}  \;;\;\;\;  5 < \tan\beta < 40 
  \label{eq:range2} 
 \end{equation}
 and  sign$(\mu)>0$. Note that in order to be more general we have not fixed the right sneutrino mass at $m_0$ but rather used its value at the electroweak scale as a free parameter. This could qualitatively be justified since sneutrinos are gauge singlets. In any case we will show the impact of this assumption.  

\begin{figure}[!htb]
\begin{center}
{
\includegraphics[width=10cm,height=10cm]{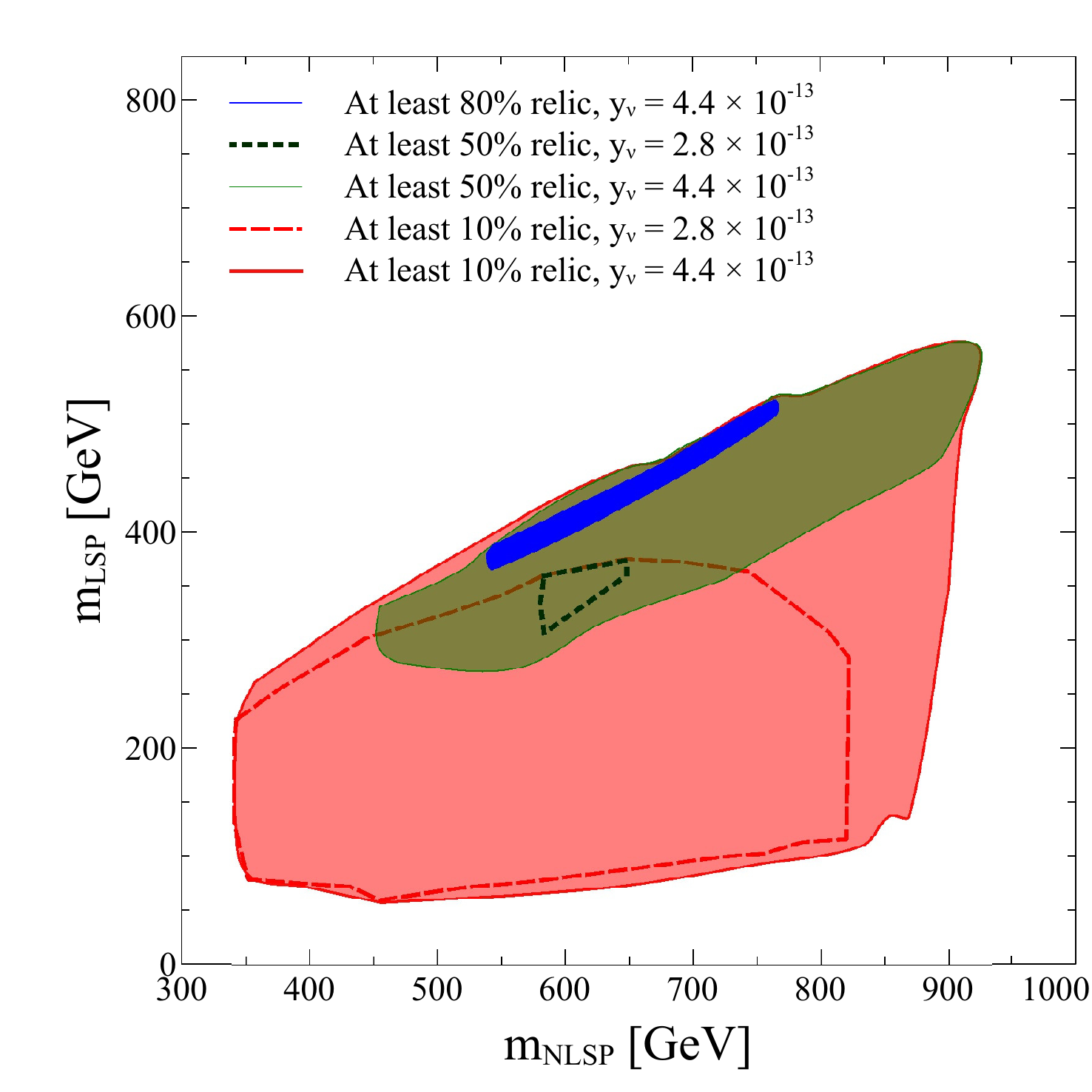}
\caption{Allowed parameter region showing percentage relic abundance in the $m_{\stau}-m_{\tilde{\nu}_R}$ 
$(m_{\textrm{NLSP}}-m_{\textrm{LSP}})$ space for two different values of the  Yukawa coupling corresponding to the  degenerate 
 and `hierarchical'  neutrino masses. }
\label{fig:DM_per}}
\end{center}
\end{figure}
 
 \begin{figure}[!htb]
\begin{center}
{
\includegraphics[width=10cm,height=10cm]{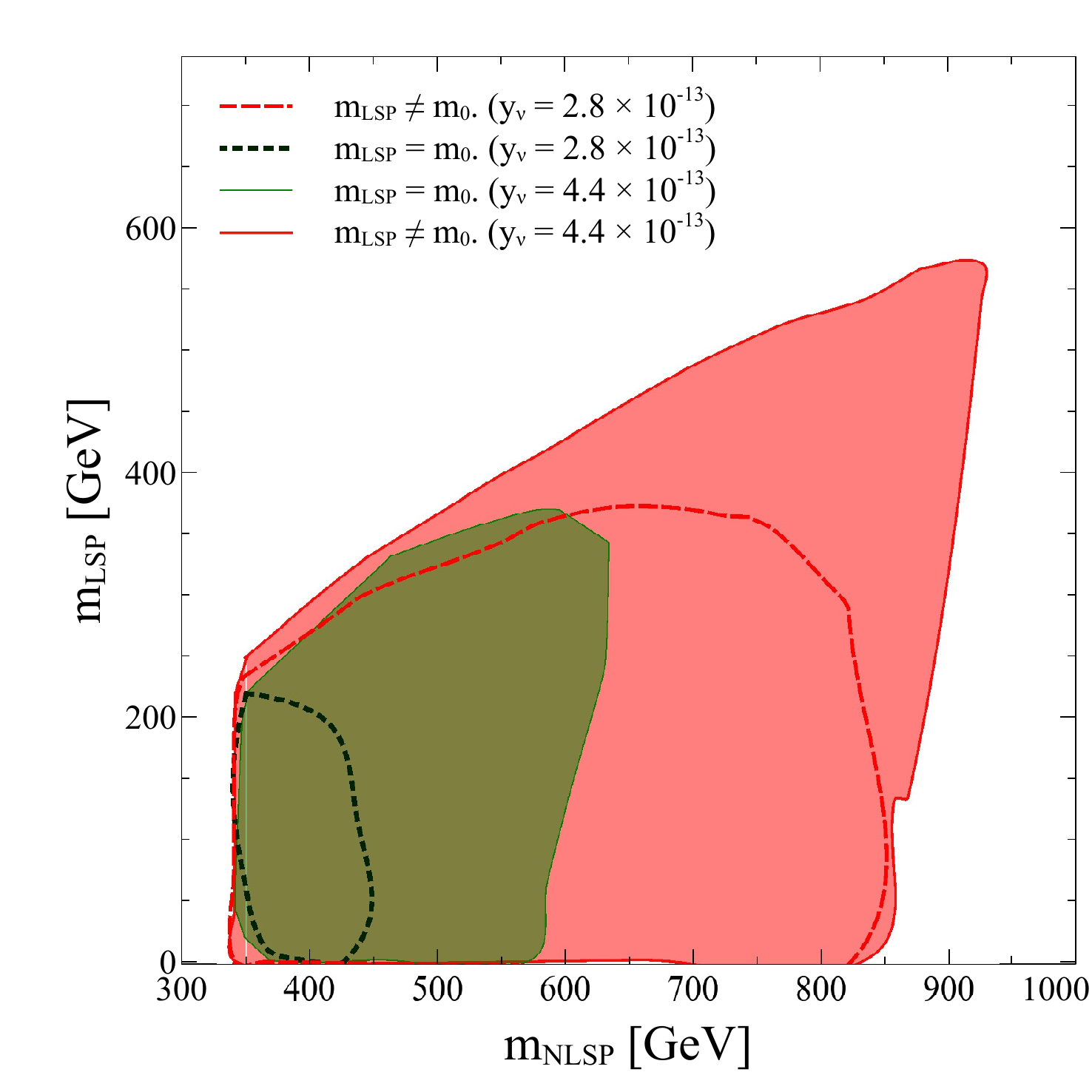}
\caption{Allowed parameter region showing the change in assuming $m_{\tilde{\nu}_R} = m_0$ in the $m_{\stau}-m_{\tilde{\nu}_R}$ 
$(m_{\textrm{NLSP}}-m_{\textrm{LSP}})$ space for the `hierarchical' (dashed) and degenerate (solid) neutrino masses. 
}
\label{fig:m0}}
\end{center}
\end{figure}

 Our benchmark case assumes quasi-degenerate neutrino masses, with $y_{\nu} = 4.4 \times 10^{-13}$.
To show the sensitivity to this value, we shall also present how the results would change for a case with
 $y_{\nu}=2.83 \times 10^{-13}$, loosely inspired to the mass scale associated to a normal hierarchy 
 pattern, but without any flavour structure being imposed. 
 In figure~\ref{fig:DM_per}, we show the allowed region in the $m_{\stau}-m_{\tilde{\nu}_R}$ parameter space, denoted $m_{\rm NLSP} - m_{\rm LSP}$.  Clearly regions exist 
 where more than 50\% of the relic abundance can be accounted for via $m_{\tilde{\nu}_R}$'s. We even find a region 
 in parameter space where more than 80\% of the relic abundance can be accounted for, which 
 loosely corresponds to accounting for the totality of the DM after including theoretical uncertainties, for example from higher order effects~\cite{Baro:2007em}.
 Note that the BBN constraint, by imposing an upper limit on the stau lifetime, effectively removes the region of small $m_{\stau}-m_{\tilde{\nu}_R}$  mass splitting where the 3-body decay of the stau dominates and even some of the region where only 2-body  decays occur.  
 Note that the allowed regions shrink when the lower value of
 the Yukawa coupling is adopted (dashed contour) and no region can be found where $m_{\tilde{\nu}_R}$'s contribute 80\% or more to the DM. This is because a smaller Yukawa means a smaller sneutrino mixing, leading to a longer lifetime and more severe BBN constraints.
 Interestingly enough, {\it the fate of this DM candidate is linked to the sharpness of the cosmological neutrino mass constraints}.
 A positive detection of the neutrino mass in the degenerate limit, \textit{i.e.} just around the corner, would be compatible with the scenario
 described in this paper accounting for most if not all of the DM. Should the neutrino mass pattern be constrained to (or detected at)  a hierarchical  spectrum,
 at most a sub-leading DM role would be possible for $m_{\tilde{\nu}_R}$'s in our scenario.

In figure~\ref{fig:m0}, we show how the parameter region changes once we impose the universality condition on the RH sneutrino mass, more precisely we demand that $|m_{{\tilde{\nu}}_R} - m_0| < 5$ GeV. 
This amounts to removing one free parameter of the model. For a given value of $m_{\tilde{\nu}_R}$   we therefore expect  a reduced upper bound on the mass of the  stau NLSP, even before imposing any constraints,  since for a fixed $m_0$  the range of predicted masses are determined only by the term proportional to $m_{1/2}$ in the RGE.
The impact of the BBN constraints is, in both scenarios,  to restrict the region at large $m_{\tilde \tau_1}$ (and  $m_{\tilde{\nu}_L}$). We found however that 
the upper bound on the mass of the NLSP is much lower ($m_{\tilde \tau_1} \approx 600$~GeV) in the restrictive (``unified'') scenario.  The reason is two-fold. First,  
the sneutrino mixing angle is suppressed for large $m_{{\tilde{\nu}}_L}$, see Eq.~(\ref{eq:mixing}); this leads to a longer lifetime and thus tighter constraints from $^2$H/H.
Moreover, we found more points with large mixing in the generic scenario than in the restrictive one. 
Secondly,  the relic density of the NLSP depends on all parameters of the stau sector, and in particular on $m_0$, as it enters into the calculation of the stau-annihilation rate.  For similar NLSP and LSP masses we found  larger values for 
$\Omega_{\tilde \tau_1}$, which implies a larger yield $Y_{NLSP}$ in the unified model, in turn leading to a more stringent constraint from $^4$He. 
As a result, heavier $\stau$'s are excluded in the more restrictive (``unified'') scenario, but as long as the Yukawa coupling is close enough to the current degenerate upper limit, the reduction of the allowed parameter space is not too dramatic.

\begin{figure}[!htb]
\begin{center}
{
\includegraphics[width=10cm,height=10cm]{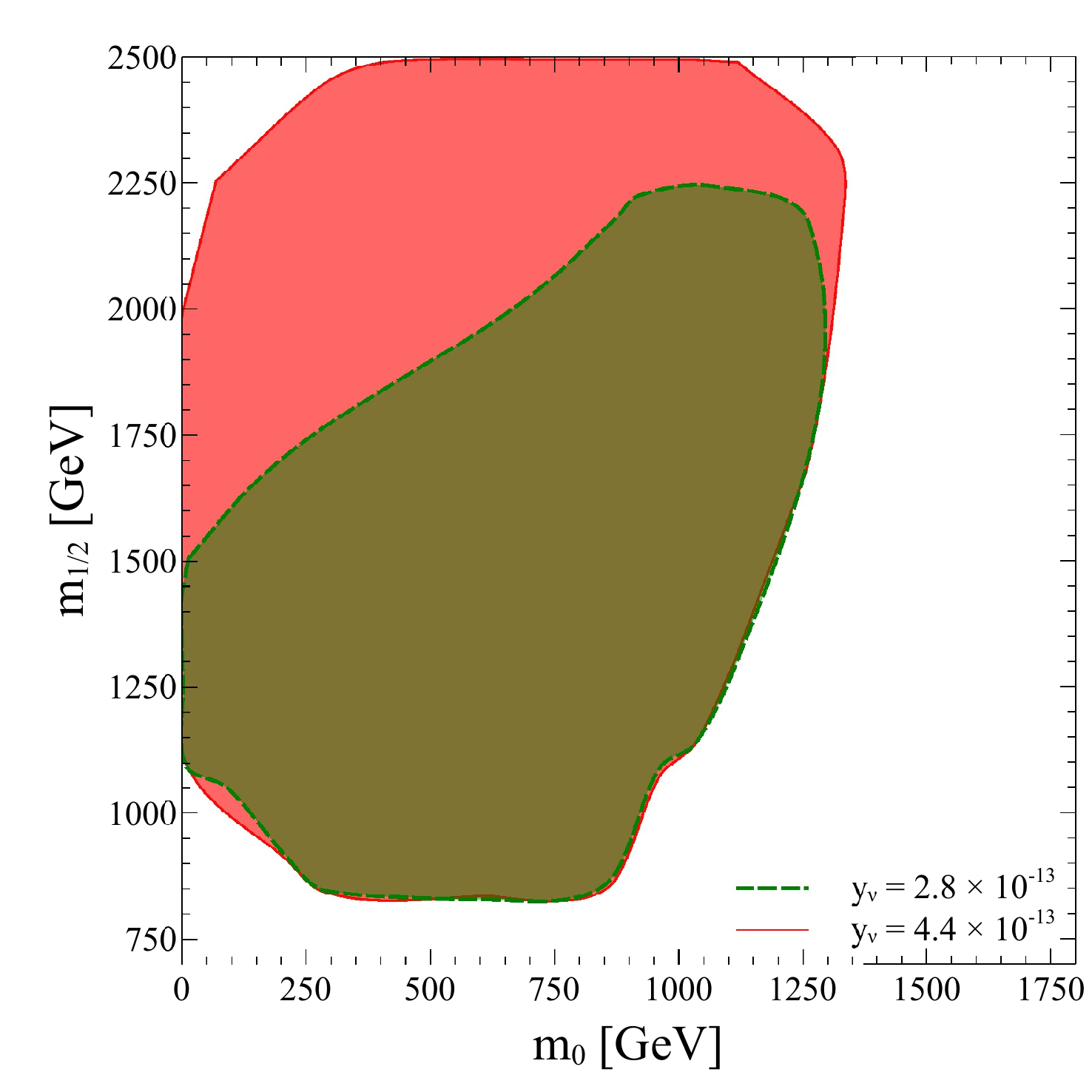}
\caption{Allowed parameter region in the $m_0-m_{1/2}$ plane which satisfies all existing collider, low energy, relic and
BBN constraints for the `hierarchical' (green) and degenerate (red) neutrino masses. Here we have scanned the parameters
as follows: $m_{0,1/2} < 2500$ GeV, $|A_0| < 3000$ GeV, $5 < \tan{\beta} < 40$, $0 < m_{\tilde{\nu}_R} < m_0$ and sign$(\mu) > 0$.}
\label{fig:m0m12}}
\end{center}
\end{figure}

To get an idea of the impact of the different bounds in another direction in parameter space, in Fig.~\ref{fig:m0m12}
we show the allowed region in the $m_0-m_{1/2}$ plane for the two values of the neutrino Yukawa coupling; here  $A_0$ and $\tan{\beta}$ vary in the range specified in 
Eq.s~(\ref{eq:range1}),~(\ref{eq:range2}) and sign$(\mu) > 0$.
Note that this allowed parameter space is  different from the one obtained in the CMSSM, see for example the result shown by ATLAS with the Run 1 data ~\cite{Aad:2015iea}  or
 ~\cite{Roszkowski:2014wqa}.
The main difference is that the region at very low $m_0$, forbidden in the CMSSM because the LSP is charged, is now re-open. Moreover in the CMSSM  the DM relic density imposes the stau and neutralino to be almost degenerate at low $m_0$ for coannihilation to take place, while in the $\tilde\nu$CMSSM  this mechanism is not required and values of $m_0$ above the TeV are allowed. 

\begin{figure}
\hspace*{-0.3cm}
\begin{center}
{
\includegraphics[width=10cm,height=10cm]{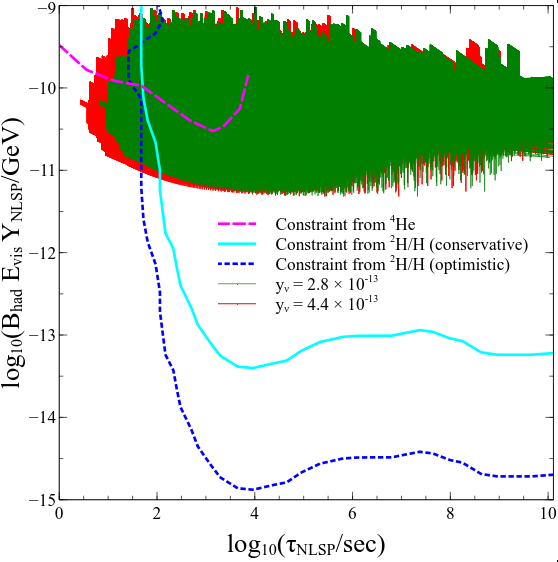}
\caption{Allowed parameter region (below the $^4$He line and to the left of the $^2$H/H line) in the lifetime-injected hadronic energy plane which 
satisfies all existing collider, low energy, relic and BBN constraints for the `hierarchical' (green) and degenerate (red) 
neutrino masses. The two curves denote the constraint from $^4$He (magenta dashed) and from  $^2$H/H (cyan solid) abundance. 
The dotted (blue) curve represents the impact of assuming a more tightening $^2$H/H determination.}
\label{fig:BBN}}
\end{center}
\end{figure}

Finally, in Fig.~\ref{fig:BBN}  we show the impact of BBN constraints from $^4$He and $^2$H/H in the relevant parameter
space, essentially determined by the lifetime of the long-lived $\stau$, denoted $\tau_{NLSP}$,  and the ``visible energy''  $E_{vis}=\frac{m_{\stau}^2+m_W^2-m_{\tilde{\nu}_R}^2}{2 m_{\stau}}$, with $B_{had} = 2/3$ corresponding to the hadronic branching fraction of $\stau$ for two body decays. The number density to entropy density ratio at $\stau$ freeze-out, 
$Y_{NLSP}$, is determined by {\tt micrOMEGAs}.

Note that the main impact of reducing the neutrino Yukawa coupling is to ``shift'' the lifetime of the $\stau$ to longer values, tightening the bounds. The viable region
is hence cornered in a small region of the cosmological parameter space, and could be further tightened by an improved neutrino mass limit, and/or an improved $^4$He
determination. Note that the $^2$H constraining power is basically saturated, with a determination more aggressive by a factor 36 only improving the lifetime bound
by $\sim 20\%$ or so.

\section{Prospects at the LHC}
\label{LHC14}
In this section we study the discovery prospects of the $\stau$-NLSP at the future runs of the LHC. We focus on the following
channels, \textit{viz.}
\begin{itemize}
 \item $2 \, \stau + N \,\textrm{hard}\, \textrm{jets}\: (N\ge 2)\,,$
 \item $2 \, \stau$ (two stable charged tracks)\,,
 \item passive detection of highly-ionizing (slow) particles,
\end{itemize}
described in Sec.~\ref{withjets}, Sec.~\ref{tracksonly}, and Sec.~\ref{moedal}, respectively.
Out of the allowed parameter region determined in the previous section, we take four benchmark points   with increasing $\stau$ mass and show
their discovery prospect at the 14 TeV run of the LHC with an integrated luminosity up to  3000 fb$^{-1}$. The four
benchmark points are listed in Table~\ref{tab:BPs}.

\begin{table}
\centering
\begin{tabular}{|c|c|c|c|c|}
\hline
Parameter  & Benchmark 1  & Benchmark 2 & Benchmark  3  & Benchmark  4 \\ 
\hline \hline
$m_0$ & $99$, & 284, &  690 &  944 \\
$m_{1/2}$  & 1048  &  961  &1369 &  1661  \\
$A_0$ &  -1897 & -2115 & -206, &  - 2175\\
$\tan{\beta}$ & 10.35 & 11.18 &  33.49 & 38.67\\
\hline
$\mu$ & 1620 & 1590 & 1923 & 2202 \\
\hline
$m_{\tilde{e}_L}, m_{\tilde{\mu}_L}$ & 705 & 701 & 1138 & 1443 \\
$m_{\tilde{e}_R}, m_{\tilde{\mu}_R}$ & 408 & 460 & 859 & 1127 \\
$m_{\tilde{\nu}_{{e}_L}}, m_{\tilde{\nu}_{{\mu}_L}}$ & 700 & 697 & 1134 & 1440 \\
$m_{\tilde{\nu}_{{\tau}_L}}$ & 687 & 679 & 1011 & 1275 \\
$m_{\stau}$ & 357 & 399 & 442 & 598 \\
$m_{\tilde{\tau}_2}$ & 694 & 687 & 1024 & 1286 \\
\hline \hline
$m_{\chi_1^0}$ & 447 & 409 & 594 & 727 \\
$m_{\chi_2^0}$ & 848 & 778 & 1121 & 1366 \\
$m_{\chi_1^{\pm}}$ & 848 & 778 & 1121 & 1366 \\
\hline \hline
$m_{\tilde{g}}$ & 2295 & 2121 & 2956 & 3543 \\
\hline \hline
$m_{\tilde{u}_L}, m_{\tilde{c}_L}$ & 2088 & 1947 & 2754 & 3321 \\
$m_{\tilde{u}_R}, m_{\tilde{c}_R}$ & 2000 & 1868 & 2642 & 3185 \\
$m_{\tilde{d}_L}, m_{\tilde{s}_L}$ & 2089 & 1948 & 2755 & 3322 \\
$m_{\tilde{d}_R}, m_{\tilde{s}_R}$ & 1991 & 1860 & 2629 & 3170 \\
$m_{\tilde{t}_1}$ & 1385 & 1210 & 1914 & 2358 \\
$m_{\tilde{t}_2}$ & 1849 & 1698 & 2351 & 2819 \\
$m_{\tilde{b}_1}$ & 1814 & 1659 & 2316 & 2783 \\
$m_{\tilde{b}_2}$ & 1970 & 1834 & 2423 & 2875 \\
\hline \hline
$m_{h^0}$ & 124  & 124 & 125 & 126 \\
$m_{A^0}$ & 1739 & 1699 & 1764 & 1924 \\
\hline 
\end{tabular}
\caption{Benchmark points for studying the discovery prospects of the $\stau$-NLSP in the $\tilde\nu$CMSSM framework with  a RH
sneutrino LSP. All the superparticle masses and dimensionful input parameters are shown in GeV. The $\tilde{\nu}_{i_R}$ masses are not  shown in the table as the exact value is not important for the collider phenomenology provided  $m_{\tilde{\nu}_{i_R}} < m_{\stau}$ .  The top mass is fixed at 173.1 GeV
and has been used for the running of the parameters.}
\label{tab:BPs}
\end{table}

For these benchmarks, the low energy spectrum follows the general trend, 
\begin{equation}
 m_{\tilde{\nu}_{R}} < m_{\stau} < m_{\chi_1^0} < m_{\tilde{e}_1, \tilde{\mu}_1} < \ldots < m_{\tilde{g}} \nonumber
\end{equation}
suggesting that all superparticle productions at the LHC would finally end up decaying to the sneutrino-LSP. However, we must
note that the lifetime of the $\stau$-NLSP varies roughly between a few seconds to a little more than three minutes for the
allowed parameter space.  Thus these particles will decay 
only outside the detector. Within the general purpose ATLAS and CMS detectors, characteristic signatures consist of charged tracks with large transverse momenta. This is
in contrast with the standard SUSY signals where the signature involves  a substantial amount of missing transverse momenta. Thus, the ``stable''
$\stau$ will behave just like a  {\it slow} muon, \textit{i.e.} their velocity $\beta = p/E$ is appreciably lower than 1,  implying that they will have high specific ionisation. Many existing studies in the literature capitalise on the ionisation properties and the time of flight measurements of
these particles and distinguishes them from the muons~\cite{Drees:1990yw,Nisati:1997gb,Martin:1998vb,Feng:1999hg,Ambrosanio:2000ik,Buchmuller:2004rq,Ellis:2006vu}. 
There is another approach of separating the signal from the
backgrounds by looking at certain kinematic distributions~\cite{Gupta:2007ui} and giving hard cuts on these. In this work, we 
compare the two approaches for the four benchmark points listed above. We also briefly mention an unconventional passive search strategy fully relying on this property.

\begin{figure}
\begin{center}
{
\includegraphics[width=12cm,height=8cm]{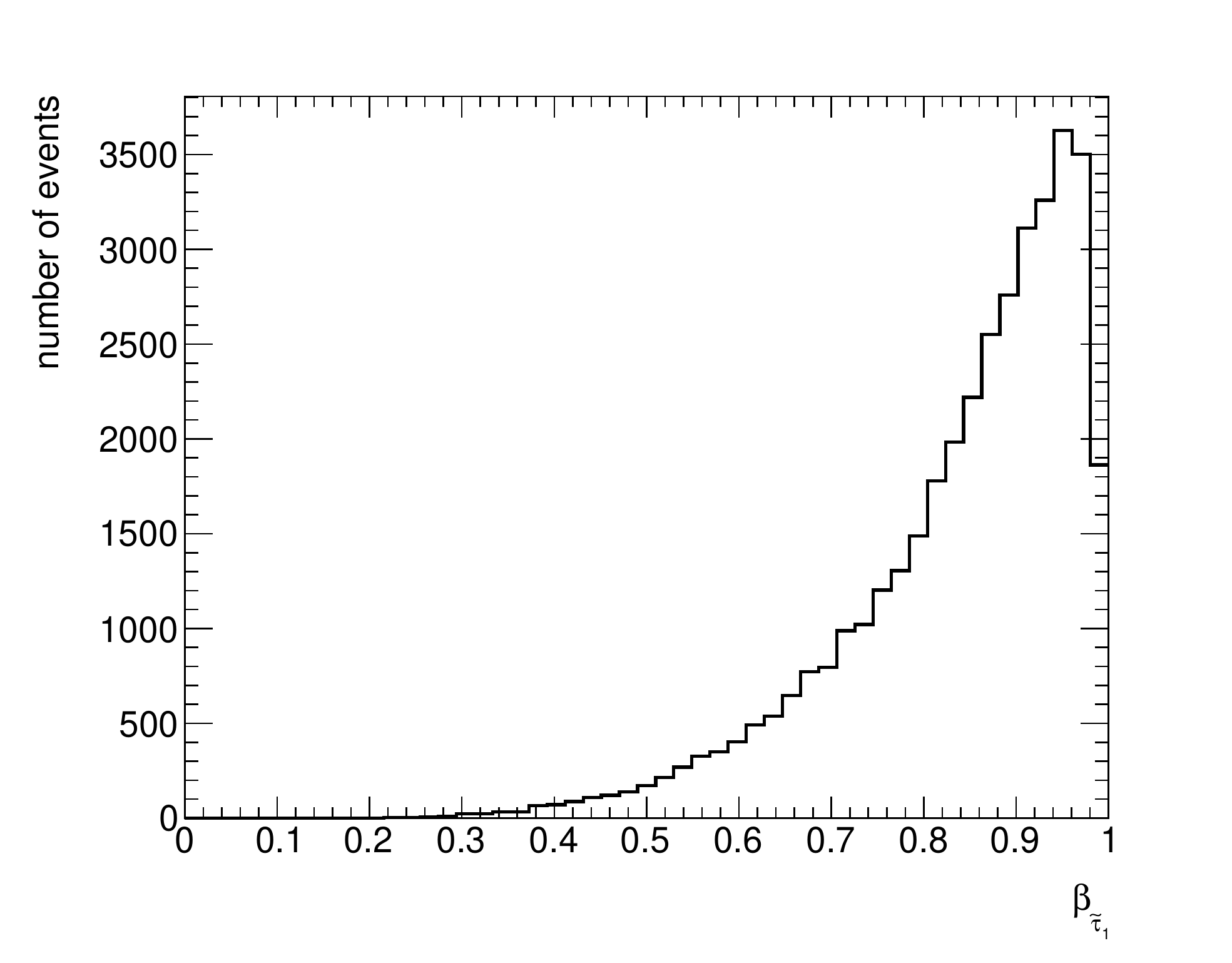}
\caption{Velocity distribution of the $\stau$-NLSP for benchmark point 3. The mean velocity is $\sim 0.84\,c$ with an root-mean-square of
$\sim 0.13\,c$.}
\label{fig:vel-dist-BP1}
}
\end{center}
\end{figure}

Before we start discussing the analysis strategy, we comment on the mass measurement strategy of these long lived charged 
particles using the time of flight measurements~\cite{Hinchliffe:1998ys}. When the staus are pair  produced,  a majority of them have a high velocity. We show the velocity distribution for the third benchmark point in Fig.~\ref{fig:vel-dist-BP1} for stau pair production. This velocity distribution can also be obtained from the time-of-flight measurement in the
muon detector system. Combining this with the measured momentum in the same system gives  the mass of the particle using the
relation
\begin{equation}
 m = p/{\beta \gamma},
\end{equation}
where $\gamma$ is the Lorentz factor. The details of this measurement technique can be found in 
Refs.~\cite{Hinchliffe:1998ys,ATLAS:2014fka}. Here, we follow a fairly simple-minded approach. Instead of
taking the mass of the $\stau$ at its fixed value obtained from the SUSY spectrum, we smear its mass as a gaussian with a standard deviation
of 5 GeV which is roughly of what is obtained in Ref.~\cite{Hinchliffe:1998ys} considering the uncertainty from the time-of-flight
measurements. In doing so, we generate a gaussian random number with its mean as the value of $m_{\stau}$ obtained from the SUSY-spectrum
and a standard deviation of 5 GeV. We use the Box-Muller transform in generating the gaussian random numbers.

It is also important to comment on the velocity distribution of the muon which is the single most important candidate for our
backgrounds. Because we consider $\beta$ to be an important observable, it is essential to obtain a realistic velocity distribution
for the muons. However this is very difficult to mimic from fast detector simulations. 
Thus, we again refer to the experiments. The velocity 
distribution of the muons from a combined measurement of the calorimeter and the muon spectrometer has a small spread with
a mean value of $\bar{\beta}= 0.999\,c$ and a standard deviation of $\sigma_{\beta} = 0.024\,c$, see Fig. 1 (right) in Ref.~\cite{ATLAS:2014fka}. 
Hence, in our analysis we generate
a gaussian random number with these parameters and then impose the cuts on $\beta$ accordingly.
We must note  in passing that in BP1  the RH selectron and
smuon states are lighter than $\chi_1^0$. In such cases, the neutralino can decay into $\tilde{e}e(\tilde\mu\mu)$. The decays of the charged slepton  to  the lighter sneutrino state of the first two families are highly suppressed  by the small Yukawa couplings, as is the case for the $\stau$-NLSP. However the competing  three body decay  via a virtual neutralino into
the $\stau \tau e(\mu)$ or $\tilde\nu_\tau \tau \nu_e(\nu_\mu)$as the final state particles will prevent the selectrons (smuons) 
from being long-lived. These can however give additional events with $\stau$-tracks. In the analysis below we will not consider the direct production of selectrons and smuons, this process was studied in~\cite{Gupta:2007ui}.

In the next two subsections, we discuss the two proposed final states and investigate the discovery prospects of the long-lived
$\stau$s. We compute the statistical significance using the standard formula
\begin{equation}
 \mathcal{S} = \frac{N_S}{\sqrt{N_S + N_B}},
\end{equation}
where $N_S$ and $N_B$ are respectively the number of signal and background events passing the selection cuts.

\subsection{Two $\stau$ and at least two hard jets}\label{withjets}
To perform our analysis, we generate the SUSY-spectra using {\tt SPheno}~\cite{Porod:2003um,Porod:2011nf}. The output {\tt SLHA}~\cite{Skands:2003cj}
files are fed into the {\tt MadGraph5 aMC@NLO}~\cite{Alwall:2014hca} program to generate the signal events. The showering and 
hadronisation is done using {\tt Pythia 6}~\cite{Sjostrand:2006za}. Finally the detector simulation is done in the 
{\tt Delphes 3}~\cite{deFavereau:2013fsa} framework~\footnote{We thank Pavel Demin, Shilpi Jain and Michele Selvaggi for technical help in
implementing the $\stau$s as stable charged tracks in the {\tt Delphes 3} framework.}. In order to decay the $\chi_1^0$ in  
{\tt Pythia}, we had to modify the main code slightly  since the lightest neutralino is by default considered to be the LSP. For the signal generation,
the parton distribution functions have been evaluated at $Q=2 m_{\stau}$ using {\tt CTEQ6L1}~\cite{Pumplin:2002vw}. The renormalisation
and factorisation scales are set as
\begin{equation}
 \mu_R = Q = \mu_F
\end{equation}
The jets have been formed using the anti-$k_t$ jet clustering algorithm~\cite{Cacciari:2008gp} in the {\tt FASTJET} 
framework~\cite{Cacciari:2011ma} with the $R$ parameter set equal to 0.6. The signal
cross-sections have been rescaled by their next-to leading order (NLO) $k$-factors using {\tt Prospino 2}~\cite{Beenakker:1999xh}.
We generate the signal samples together in {\tt MadGraph} which also gives  the cross-sections in the separate
channels like squark-squark, squark-gluino and gluino-gluino production. We then compute the $k$-factors for each of these using 
{\tt Prospino} and computed the effective $k$-factor by weighting with the cross-sections in the individual channels.

\begin{figure}
\begin{center}
{
\includegraphics[width=7.5cm,height=5.5cm]{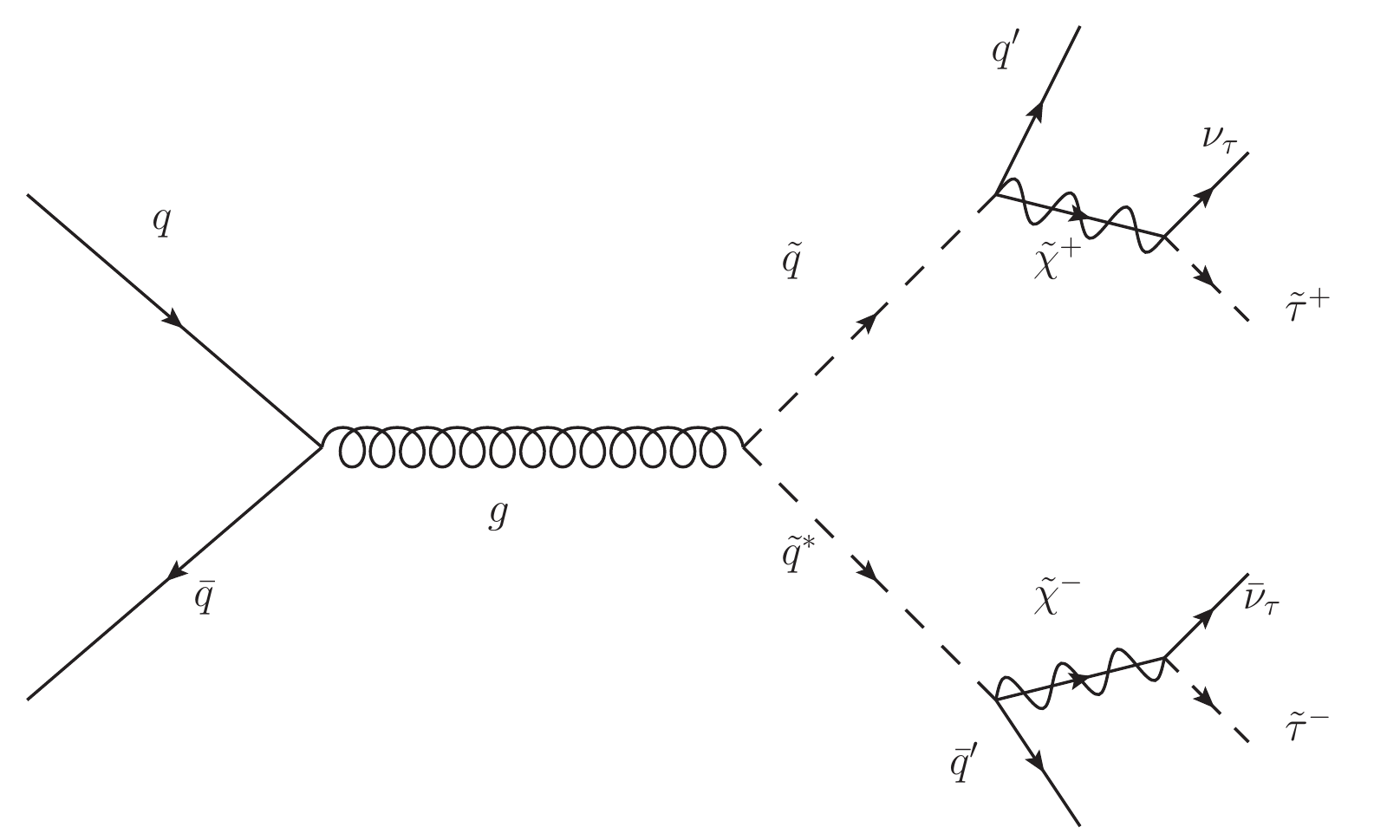}~~
\includegraphics[width=7.5cm,height=5.5cm]{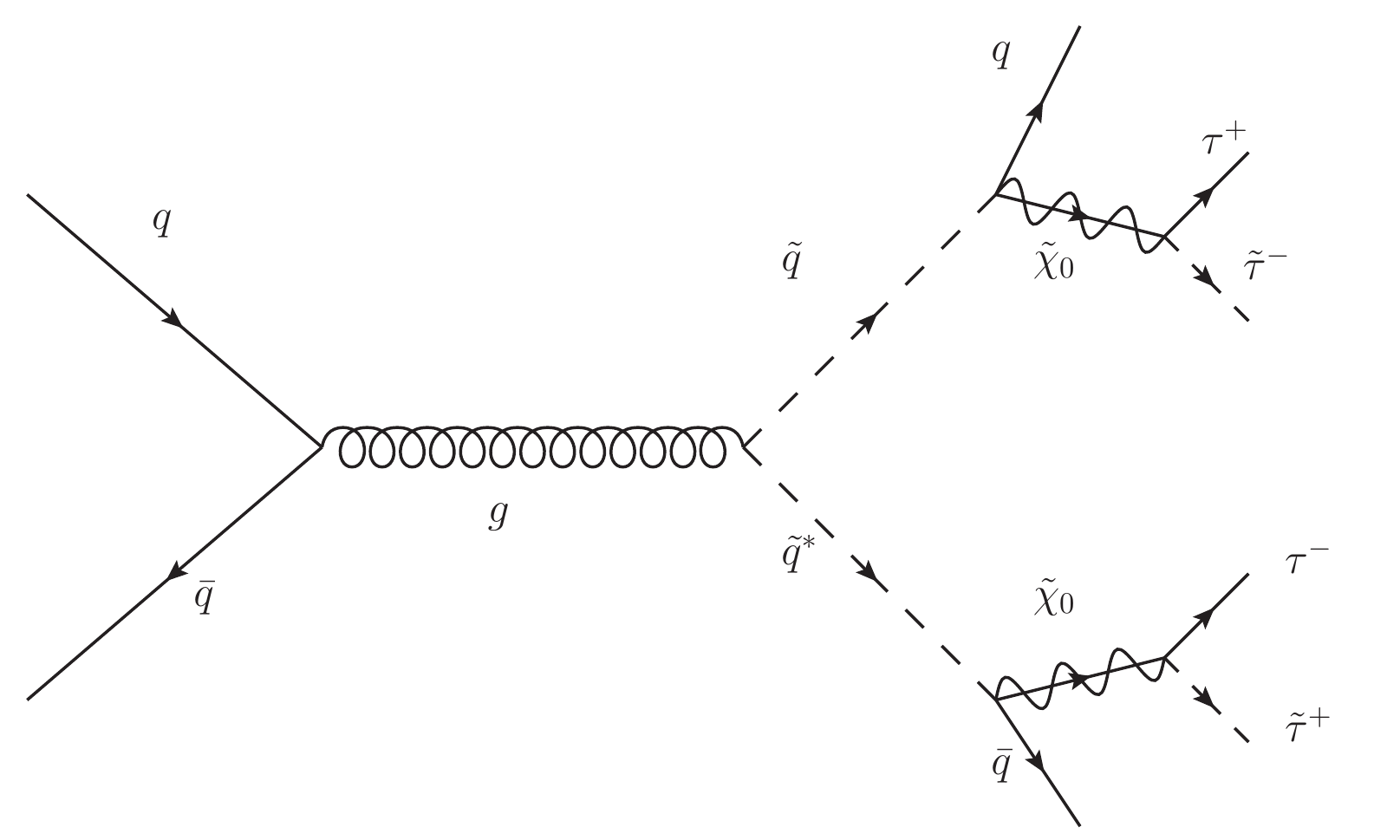}
\caption{Sample Feynman diagrams showing a pair production of $\stau$s with additional jets and $\slashed{E}_T$ or jets and leptons.}
\label{fig:FeynDiag}
}
\end{center}
\end{figure}

Since  stable staus appear in the SUSY decay chains, the main contribution to this channel comes from squark pair production (sample Feynman diagrams are shown in Fig.~\ref{fig:FeynDiag}). The production of one or two gluinos are suppressed relative to the squarks due to the higher mass of the gluino, see Table~\ref{tab:BPs}, and the fact that for such high masses the gluon PDF's are small. 
For this channel the dominant backgrounds are : $t\bar{t}$ (computed at N$^3$LO~\cite{Muselli:2015kba}) and the Drell Yan 
production of $\mu \mu (\tau \tau) +$ jets (computed  at NNLO~\cite{Catani:2009sm}). For the latter, we take a matched sample, matched
up to 3 jets using the MLM ME-PS matching scheme~\cite{Alwall:2007fs}. Besides these, the other contributions to the backgrounds come from
$W^+W^-$, $WZ$ and $ZZ$  and are computed at NLO~\cite{Campbell:2011bn}. Here we take a similar approach as considered in Ref.~\cite{Gupta:2007ui}.
Instead of considering the velocity of the stable staus, we consider hard kinematical cuts, specifically 

\begin{itemize}
 \item $p_T^{\mu_{1,2}} > 200$ GeV, $|y(\mu_{1,2})| < 2.4$,
 \item $p_T^{j_{1,2}} > 200$ GeV, $|\eta(j_{1,2})| < 5.0$,
 \item $\sum{|p_T^{vis.}|} > 1000$ GeV,
 \item $\Delta R(\mu_1,\mu_2) > 0.2$,
 \item $\Delta R(j, j) > 0.4$,
 \item $\Delta R(\mu, j) > 0.4$,
 \item $M_{\mu_1, \mu_2} > 1000$ GeV,
\end{itemize}
where the subscripts 1 and 2 refer to the hardest or the second hardest object when ordered by $p_T$, and  $p_T^{\mu}$ generically refers to the track, be it due to the $\stau$ signal or the background muons.
These cuts have a  dramatic effect in removing the backgrounds almost completely. To perform the analysis, we generated a statistically significant number of
events such that we are sure of the number of events after the cuts.
With the above cuts, we end up with 0 events for $WZ+$ jets and $ZZ+$ jets. In Table~\ref{tab:sig-cascades} we show  the luminosity required to reach a 
5$\,\sigma$ statistical significance significance  for stable staus for each of the benchmarks. The small number of background events surviving the cuts scales with the luminosity considered for each point. 
Note that the lower luminosity required for BP2 as compared to BP1  is linked to the fact that the coloured state are lighter for BP2.

\begin{table}[!ht]
\centering
\begin{tabular}{|c|c|c|c|c|c|}
\hline 
Benchmark point & $\mathcal{L}$ for 5$\sigma$ $[\textrm{fb}^{-1}]$ & $N_S$ & $N_B$ & $N_S/N_B$  \\
\hline
\hline
BP1             & 9.10                                              & 25.26 & 0.35  & 72.17   \\ 
BP2             & 2.45                                              & 25.19 & 0.09  & 265.2 \\
BP3             & 68.50                                             & 27.42 & 2.67  & 10.27  \\
BP4             & 1100                                              & 47.59 & 42.87 & 1.11   \\
\hline
\end{tabular} 
\caption{The luminosity required in order to attain a 5$\sigma$ statistical significance for stable staus for the four benchmarks. The 
number of signal and background events  as well as the ratio $N_S/N_B$ after the selection cuts for that  particular luminosity is also displayed. }
\label{tab:sig-cascades}
\end{table}

From these results  we conclude that it is fairly simple to probe collider stable staus with masses $\lesssim 400$ GeV (BP1 and BP2) in the early runs
of the 14 TeV LHC. Moreover, we estimate with a simple rescaling of the LO cross sections from 14TeV to 13 TeV,  that a  5$\,\sigma$ significance is also reachable with the 13 TeV Run with luminosities of roughly 15 (4)fb$^{-1}$ for BP1 (BP2). For larger stau masses (linked with heavier coloured states in our model) one gradually requires larger integrated luminosity: $\mathcal{L}=1000 \textrm{ fb}^{-1}$ allows to probe masses up to $580\,$GeV , with the full integrated luminosity  $\mathcal{L}=3000 \textrm{ fb}^{-1}$, the mass reach can be  extended to roughly $600\,$GeV, thus allowing to cover a significant fraction of the currently allowed parameter space.

\subsection{Two $\stau$ tracks}\label{tracksonly}

Here, we study the discovery prospects  of  directly  produced $\stau$ pairs. This channel suffers from a smaller production cross section as compared to the previous one (electroweak production as compared to strong production) but presents the advantage of being  fairly model
independent: This channel can also be used beyond the $\tilde\nu$CMSSM framework when the coloured states are too heavy to be produced at a significant rate. 
Moreover  we can directly use the constraints already set by CMS and ATLAS from the run I data, i.e. a lower bound
on the $\stau$ mass of $289$ GeV and $340$ GeV from ATLAS~\cite{ATLAS:2014fka} and CMS~\cite{CMS-PAS-EXO-15-010,Chatrchyan:2013oca}, 
respectively. These bounds are from the tracker plus time-of-flight measurements. CMS quotes a much relaxed lower bound on
the $\stau$ mass at 190 GeV from the tracker measurement alone. CMS has also quoted the bound from the 13 TeV run and it is
weaker than its 8 TeV counterpart, \textit{viz.} $m_{\stau} > 230$ GeV. However, we do not consider this bound
and take the more stringent one because the 13 TeV run till now has a significantly small integrated luminosity.

Let us also stress  that we have used the Drell-Yan plus $b\bar{b}$ initiated production for the $\stau$ pairs in obtaining
the final results. It is mentioned in Ref.~\cite{Lindert:2011td} that the gluon fusion initiated processes can enhance the cross-sections
by an order of magnitude. However, this statement holds for $\stau$ masses below 250 GeV, with the maximum effect achieved when 
100 GeV $< m_{\stau} <$ 200 GeV. However, in our case the stau is always heavier than 340 GeV. We find that with 
an increase in the masses of the particles in the loop, \textit{viz.} the squarks, we have decoupling and this does not 
lead to any contribution from such loops: we checked explicitly by using MadGraph5 aMC@NLO that at LO, the gluon initiated processes have contributions of the order of $\mathcal{O}(10^{-9}-10^{-8})$ fb.

For this particular channel the most dominant background is a pair of muons.  We generate pairs of muons 
as well as pairs of  taus (which can also lead to a two-muons final state) using MadGraph5 aMC@NLO
and use the same procedure for the detector analysis as for the signal. Here we also fold in 
 the next-to-next-to leading order (NNLO) $k$-factor~\cite{Catani:2009sm}. We also consider the subdominant
 backgrounds, \textit{viz.} $t\bar{t}$ and diboson pairs ($WW,WZ$ and $ZZ$) computed respectively at N$^3$LO~\cite{Muselli:2015kba} 
 and NLO~\cite{Campbell:2011bn}.~\footnote{For low $p_T$ cuts ($p_T \sim 15$ GeV), muons from $b$- and $c$-decays can have  
 substantial rates~\cite{Heisig:2011dr}. But because we impose a high $p_T$ cut on the muons (see Table~\ref{tab:cuts}), and we also require a jet-muon isolation 
 cut, these backgrounds become negligible. Hence we do not explicitly include these backgrounds in our analysis.} We use the following basic trigger cuts for our 
analysis,
\begin{itemize}
 \item $p_T^{\mu} > 70$ GeV,
 \item $|\eta(\mu)| < 2.5$,
 \item $\Delta R(\mu \mu) > 0.4$.
\end{itemize}
These same cuts have been applied to the $\stau$ tracks in the detector analysis because of our inability to generate the samples
with such trigger cuts at the generator level using MadGraph. After this we used three sets of selection cuts to see which 
fares better in terms of the significance. In table~\ref{tab:cuts}, we list down the selection cuts in details. The cut set C
resembles the one used by ATLAS~\cite{ATLAS:2014fka}.

\begin{table}
\centering
\begin{tabular}{|c|c|c|c|}
\hline
Cut on  & Cut set A  & Cut set B & Cut set C  \\ 
\hline \hline
$\beta$ &  $> 0.85$ & $-$ & $ < 0.95$ \\
$p_T^{\mu_{1,2}}$ & $> 200$ GeV & $> 200$ GeV & $> 70$ GeV \\
$\sum{|p_T^{vis.}|}$ & $ > 700$ GeV & $ > 500$ GeV & $-$ \\
$|y(\mu_{1,2})|$ & $ < 2.4$ & $ < 2.4$ & $ < 2.5$ \\
$M_{\mu_1, \mu_2}$ & $> 1200$ GeV & $ > 1000$ GeV & $-$ \\ 
$\Delta R(\mu_1,\mu_2)$ & $> 0.2$ & $> 0.2$ & $-$ \\
$\Delta R(\mu,j)$ & $> 0.4$ & $> 0.4$ & $-$ \\
$\Delta R(j,j)$ & $> 0.4$ & $> 0.4$ & $-$ \\
\hline 
\end{tabular}
\caption{The three sets of selection cuts applied in the 
$\stau$ pair analysis. Set C resembles the set of cuts of the ATLAS analysis~\cite{ATLAS:2014fka}.}
\label{tab:cuts}
\end{table}

The hard $p_T$ cuts are extremely efficient in removing a significant amount of the backgrounds. However,  the Cut set C proves to be one of the most efficient ones because the muon velocity distribution peaks
roughly around unity with a very small spread~\footnote{Note that Ref.~\cite{Heisig:2011dr} uses slightly different
sets of selection cuts including $0.6 < \beta < 0.9$. We have restricted ourselves to weaker cuts on $\beta$.}.
The number of signal and background events and the significances for an integrated luminosity of 3000 fb$^{-1}$ are listed in Table~\ref{tab:sig}, showing that more than 5$\,\sigma$ significance can be reached for all points but  BP4 with Cut set C or B. 
With Set A, especially due to  the lower bound on $\beta$ and also to the more stringent cuts on $p_T$ and on the invariant mass,  
a much  larger fraction of the signal is suppressed, leading to smaller significance  even though the background is also more suppressed. 
 Note in addition that the reach on $m_{\tilde{\tau}}$ could be extended  by requiring a tighter  cut on $\beta$. Choosing  $\beta < 0.8$  as per CMS [60,61], 
  we get statistical significances of 4.7$\sigma$ and 3.0$\sigma$  for $m_{\tilde{\tau}} = 700, 800$ GeV respectively. 
This hard cut on $\beta$ renders the background vanishingly small: even by including the spread in the muon $\beta$ distribution from ATLAS [62], we
hardly get any background events which pass $\beta < 0.8$.

\begin{table}[!ht]
\centering
\begin{tabular}{|c|c|c|c|c|c|}
\hline 
Cut set   & Benchmark point &  $N_S$ & $N_B$ & $N_S/N_B$ & $\mathcal{S}$ \\
\hline 
\multirow{3}{0.5cm}{A}  & BP1 & 526 &      & 0.09 & 6.7 \\ 
                        & BP2 & 358 & 5684 & 0.06 & 4.6 \\
                        & BP3 & 258 &      & 0.05 & 3.3 \\
                        & BP4 & 47  &      & 0.01 & 0.6 \\
\hline
\multirow{3}{0.5cm}{B}  & BP1 & 1337 &       & 0.10 & 11.3 \\ 
                        & BP2 & 1069 & 12772 & 0.08 & 8.9 \\
                        & BP3 & 826  &       & 0.06 & 7.0 \\
                        & BP4 & 232  &       & 0.02 & 2.0 \\
\hline
\multirow{3}{0.5cm}{C}  & BP1 & 1543 &      & 0.44 & 21.8 \\ 
                        & BP2 & 1014 & 3481 & 0.29 & 15.1 \\
                        & BP3 & 715  &      & 0.21 & 11.0 \\
                        & BP4 & 211  &      & 0.06 & 3.5  \\ 
\hline
\end{tabular} 
\caption{Table showing the number of signal and background events after the selection cuts for the three sets of 
selection cuts, the ratio $N_S/N_B$ and the statistical significance $\mathcal{S}$. The integrated luminosity used to compute
these numbers is 3000 fb$^{-1}$.}
\label{tab:sig}
\end{table}

\subsection{Passive highly-ionizing track detection}\label{moedal}
An unconventional search strategy is also possible at the new and largely passive detector MoEDAL~\cite{Acharya:2014nyr}, mostly comprised of an array of nuclear track detector stacks surrounding the intersection region at Point 8 on the LHC ring, which is sensitive to highly-ionizing particles (a further
trapping array is only suitable for very long-lived particle, beyond the regime of interest of our model).  
This search does not require any trigger and in principle even one detected event would be enough, albeit multiple events 
would be needed for a robust discovery.  The only major condition for sensitivity to the produced $\stau$ is that the ionizing 
particle has a velocity $\beta \leq 0.2$. For illustrative purposes we report in Tab.~\ref{tab:moedal} the numbers of events with 
$\beta \leq 0.2$ expected for $\mathcal{L} = 3000$ fb$^{-1}$ after imposing that the track has $p^{\mu}_T > 5$ GeV. More 
detailed full detector simulations are required to be more quantitative but it is already clear that when the staus are 
produced from decays of coloured particles, there exists a possibility of an independent discovery via this complementary 
channel for all our benchmarks.
\begin{table}[!ht]
\centering
\begin{tabular}{|c|c|c|}
\hline 
Benchmark point & Cascade (Sec.~\ref{withjets}) & Direct (Sec.~\ref{tracksonly})  \\
\hline
\hline
BP1             & 45               & 2.5 \\ 
BP2             & 296              & 1.5 \\
BP3             & 24               & 1.1 \\
BP4             & 6                & 0.5 \\
\hline
\end{tabular} 
\caption{Number of $\stau$'s with $\beta\leq 0.2$ potentially detectable by MoEDAL
assuming an integrated luminosity of $\mathcal{L}=3000\, \textrm{fb}^{-1}$, for the four benchmarks and the two production mechanisms discussed in the text.}
\label{tab:moedal}
\end{table}

\section{Summary and conclusions}
\label{conclusions}
In this study, we have discussed the prospects of the revival of the CMSSM through one of the simplest extensions, \textit{viz.}
through the addition of three families of right handed (s)neutrinos, the $\tilde\nu$CMSSM. We showed that in a sufficient portion of the parameter space
the right handed sneutrino might become the LSP and by contributing to the relic abundance it can become a potential cold dark matter
candidate. In our study, we focused on an $R$-parity conserving scenario where the $\stau$-NLSP can be long lived, such 
that its decay occurs well after its freeze-out. We further imposed all the available constraints, from the Higgs and flavour sector, from SUSY searches at colliders, from neutrino masses and most importantly  the BBN constraints on the elemental abundance of $^4$He and  $^2$H. The latter
is particularly important in excluding virtually all parameters leading to stau lifetimes beyond a few minutes, which would alter too much the
primordial yields via the cascades induced by the $\stau$ decay byproducts.  After imposing all these constraints, one is left with regions where $m_0$ and $m_{1/2}$ 
can range up to 1.4 TeV and $\sim$2.5 TeV respectively. One also sees that on demanding more contribution to the relic, the
parameter region shrinks. Only a very narrow region remains when demanding that the sneutrino contributes to more than $80\%$ of the dark matter, and this is also
very sensitive to the actual value of the neutrino mass scale. For instance, a non-degenerate neutrino mass spectrum, either in the case of normal or inverted mass
hierarchy, would imply that the sneutrinos can at most contribute at a subleading level, ${\cal O}$(10\%), to the relic abundance.

Finally we study the prospects of observing such long-lived staus at the recent/future runs of the LHC 
with three strategies, \textit{viz.}
\begin{itemize}
 \item From the cascade decays of the production of squark pairs, gluino pairs and pairs of squark and gluino, we find 
       significant mismatch of the kinematic distributions of the signal and the backgrounds. We apply
       hard cuts on the $p_T$ of the stable tracks (which fake ``heavy'' muons) and the jets. Furthermore, from the time
       of flight measurements and the measurement of the velocity of such particles, one can indirectly measure the masses of
       the staus.
       We find that a hard cut on the total visible transverse momentum and on the invariant mass of the pair of stable tracks
       kills all the backgrounds. It is possible to observe such a long-lived stau of mass around
       400 GeV at 5$\,\sigma$ from the 13 TeV run with an approximate integrated luminosity of 4 fb$^{-1}$ and that a high luminosity run at 14 TeV would probe stable staus as heavy as 600 GeV.  The only drawback of this otherwise promising search is its model dependence, since it relies on the mass of the gluino and the squarks.
       
 \item We further show the discovery prospects of a stau when they are directly pair produced. We find that one needs much
       higher luminosities to potentially discover the staus from this channel, which however presents the advantage of being fairly model independent. With the current set of cuts from ATLAS, a stau of mass around 400 GeV can be discovered with a 14 TeV run at 300 fb$^{-1}$. 
         
 \item We briefly discussed the perspective of an additional discovery opportunity at the unconventional MoEDAL passive detector, sensitive to highly-ionizing (slow) tracks. Despite the fact that no detailed simulations are currently available, the number of slow events expected in the high-luminosity run is encouragingly high that a discovery should be within reach in a significant fraction of the parameter space.
 \end{itemize}

To conclude, let us express a few general remarks. It is interesting that a minor modification to the CMSSM (in this case demanded by the empirical evidence for 
neutrino masses) leads to major phenomenological changes and hence in the appropriate search strategies at colliders. Even though supersymmetry is being pushed to
the backseat by every new experimental set of data from the LHC, this qualitative lesson may stay true and apply to a number of alternative scenarios, when moving
beyond minimal models. This certainly motivates one to pursue further in devising more involved search techniques like the one sketched in our study.
A second consideration concerns the deep links existing between neutrino physics, early universe cosmology (dark matter, BBN) and collider searches:
the $\tilde\nu$CMSSM model discussed here is a remarkable illustration of these tight relations, to the point that for instance a neutrino mass measurement at a factor two below
current upper limits would rule out a dominant DM role of our sneutrino candidates. Finally, there are other interesting perspectives concerning cosmological and astrophysical consequence of such a kind of DM candidate. One further possibility for diagnostics relies on the fact that sneutrinos are not really ``cold'' dark matter candidates: Due to the recoil acquired in the decay they may have a sizable kinetic energy. This possibly leads to other observables, like a suppression of small scale cosmic structures due
to their free-streaming. Several of these consequences have been explored for other ``superWIMP'' candidates (see~\cite{Feng:2003xh}  and ref.s to it) and we will not repeat them here. 
On the other hand, our DM candidate has a very peculiar feature, being possibly constituted by a mixture of three almost degenerate sneutrino states. Two of the three states are very long-lived but can decay e.g. via $\tilde \nu_2\to \tilde \nu_1 \nu_\alpha\bar\nu_\alpha$. We have not explored here the associated phenomenology, since it crucially depends on the mass matrix
pattern of the right-handed sneutrinos. It is possible that some interesting cosmological or astrophysical consequences may follow. This is a further point deserving investigation, notably if forthcoming data should comfort the viability of the model discussed in this article.


\acknowledgments
We thank Bobby Acharya and James Pinfold for useful exchange on the MoEDAL potential. We thank Daniele Barducci, Sanjoy Biswas,
Arghya Choudhury, Pavel Demin, Diego Guadagnoli, Shilpi Jain, Olivier Mattelaer, Subhadeep Mondal, Satyanarayan Mukhopadhyay, Emanuele Re and Michele Selvaggi for helpful 
discussions or technical help at various stages of this work. This work is supported by the ``Investissements d'avenir, Labex ENIGMASS'', by the French ANR, 
Project DMAstro-LHC, ANR-12-BS05-006 and the Indo French LIA THEP (Theoretical High Energy Physics) 
of the CNRS. The work of BM is partially supported by funding available from the Department of Atomic Energy, Government of India, for the Regional
Centre for Accelerator-based Particle Physics (RECAPP), Harish-Chandra Research Institute.

\bibliographystyle{unsrtnat}
\bibliography{rsnu}

\end{document}